# Using Visual Aids as a Motivational Tool in Enhancing Students' Interest in Reading Literary Texts

MELOR MD. YUNUS, HADI SALEHI, DEXTER SIGAN ANAK JOHN
Faculty of Education
Universiti Kebangsaan Malaysia
43600 UKM Bangi, Selango,
MALAYSIA
melor@ukm.my; hadisalehi1358@yahoo.com; altruist_boyz@yahoo.com

*Abstract:-* This study aims to investigate the teachers' perceptions on the use of visual aids (e.g., animation videos, pictures, films and projectors) as a motivational tool in enhancing students' interest in reading literary texts. To achieve the aim of the study, the mixed-method approach was used to collect the required data. Therefore, 52 English teachers from seven national secondary schools in Kapit, Sarawak, Malaysia were selected. Five of the respondents were also randomly selected for the interview. The analysis of the data indicated that the majority of the teachers had positive perceptions of the use of visual aids. The use of visual aids enable the teachers to engage their students closely with the literary texts despite of being able to facilitate students of different English proficiency level in reading the texts with interest. This aspect is vital as literature helps to generate students' creative and critical thinking skills. Although the teachers had positive attitudes towards the use of visual aids, the study suggests that it will be more interesting and precise if it includes students' perceptions as well.

## 1 Introduction

In the Malaysian English context, where English is officially stated and taught as a second language, learning literature in English is not easy. This difficulty arises because learning literature in English as a second language (ESL) class always poses many language and cultural obstacles [1]. Thus, since literature in English has made a serious comeback into the Malaysian classroom, it becomes the foremost task for the teachers to devise lessons creatively and innovatively in order to engage students' interest in literary learning and reading.

Today's students and classrooms are becoming more diverse and unique each day. The emergence of information and communication technology (ICT) has made it possible for teachers and students to collaborate with each other in diverse ways [2]. In literature classroom, students often encounter problems in reading and understanding the assigned literary texts in the literature component. It has been proven that secondary school students do not feel motivated to read literary texts due to lack of language proficiency and inadequate supply of teaching materials [3-4]. Thus, there is a need to insert the use of visual aids in teaching literature in order to trigger students' motivation in reading literary texts. The main objective of this study is to analyze secondary school teachers' perceptions on the use of visual aids as a motivational tool in cultivating students' interest in reading literary texts used in the English syllabus. The fundamental questions addressed in this study are:

1. What are teachers' perceptions towards the use of visual aids in teaching literary texts?
2. How can the use of visual aids be integrated in teaching literary texts?
3. How can the use of visual aids be fully utilized as an effective learning tool in motivating students to read literary texts?

## 2 Benefits of Using Visual Aids in Teaching Literature

There are a number of benefits in using visual aids in teaching literature. These benefits are of paramount importance in maintaining a good momentum of reading interest among students. Firstly, using visual aids in teaching literature creates strong engagement between students and the texts [5-6]. The use of visual aids like pictures, videos and projectors encourage students to read texts with interest, which make it easier for them to understand the abstract ideas in the texts [5]. This proves the use of visual aids act as 'vehicles' that can be used to enrich and enhance the act of reading [6]. Similarly, in research related to the use of multimedia applications for language teaching and learning literature, it has been stated that the insertion of visual aids in teaching enables authentic communication between students and literary texts [7]. Thus, it allows the students to have full concentration on the texts which leads to their understandability of the story and flow of the texts.

Secondly, students will be more motivated in learning literature. The students have no interest in learning literature due to the difficulties they encounter in the literary texts even though they show positive attitude towards the implementation of literature in English syllabus [8]. In a study conducted by Sivapalan, Wan Fatimah Ahmad & Nur Khairun Ishak (2009), the importance of audio visual and other visual aids in enhancing students' interest in learning literature was shown [9]. For example, the use of voice clips in teaching poems is vital because it could help students to understand the meaning and the message of the poems better. Apart from that, the insertion of animation of texts and graphics in the web-based multimedia as a tool to teach literature increases students' interest in learning and reading literary texts.

Thirdly, the use of visual aids in literature teaching helps students to comprehend with literary concepts. Learning literature in English for ESL learners is quite challenging. Learners often encounter difficulty in understanding the literary concepts of the texts [10]. In order to cope with this learning problem, visual aids such as graphics, illustrations, pictures, audio, and video can be used to serve as a helpful tool in facilitating students' understanding of the literary concepts occurred in the texts [11]. The characteristics of the visual aids like sound, light and color can trigger and stimulate students' understanding of the texts. Furthermore, learning literature through film puts struggling readers at less of a disadvantage. It was found out that the use of films help students to visualize clearly the literary theory and cultural aspects found in the texts [12].

## 3 Method

This study used a mix-method approach in which the researcher simultaneously collected both quantitative and qualitative data using a questionnaire and face-to-face interview. The questionnaire respondents consisted of 52 English language teachers from seven national secondary schools in Kapit, the largest district in Sarawak, Malaysia. The background of these teachers ranged from novice to experienced teachers and from major to non-major in the area of English language teaching. The selection of the respondents was based upon convenience sampling. The questionnaire items were adopted from Subramaniam, Shahizah Ismail & Koo's (2003) study and Hwang and Amin Embi's (2007) study [13-14]. Moreover, in order to obtain specific information, semi-structured interview was applied in the one-on-one interview with five respondents. The interview was guided by a list of questions that were based upon the three research questions.

The first drafts of the questionnaire and the interview protocol were examined by two experts in the field of TESL and ICT to ensure the validity of the instruments. Moreover, to test the reliability of the instruments, they were piloted on three ESL pre-service teachers. Cronbach's Alpha coefficient was calculated ($r = 0.70$) and showed that the questionnaire was reliable. The respondents' comments were also taken into account for its improvement. The respondents also had no difficulty to answer the interview questions because they were clear and understandable. However, some changes were made in the questions for better understanding.

## 4 Findings and Discussion

This part discusses the general background of the respondents and analysis of the teachers' perceptions on the use of visual aids.

### 4.1 General Background of the Respondents

The respondents were varied in terms of qualification, major of study, teaching experience, literature training and English proficiency level. These varieties are actually of utmost importance as they ensure a representative balance of respondents' perception on the use of visual aids as a motivational tool in boosting students' interest in reading literary texts.

### 4.2 Teachers' Perceptions on the Use of Visual Aids

The analysis of the results showed that the majority (96.2 %) of the teachers were in agreement that the use of visual aids is relevant and enjoyable. This was probably because the use of visual aids makes it easier for the students to understand the abstract ideas in the texts. This finding was supported by the interview data in which a teacher stated that:

*Visual aid is a relevant method to enhance the students' understanding in literary texts. So it is a very good method that could be applied by all the teachers.*

Majority of the teachers (96.2%) also demonstrated positive response towards the statement 'I enjoy teaching literary texts using visual aids'. These teachers probably had sufficient knowledge on the use of visual aids and realized its importance in attracting those students who were uninterested in reading literary texts to fully participate in literature lesson. The teachers also believed that the use of visual aids could improve students' performance. It was proved when most of the teachers (94.2 %) were in favor with the statement 'I do think the use of visual aids in teaching literary texts can improve students' performance'. Even one teacher believed that:

*Visual aids are really helpful especially in teaching literature and really help low level proficiency students.*

The majority (92.3 %) of the teachers were in agreement with the statement 'I find it easier to teach literary texts using visual aids'. Since the literary texts are often abstract, it made it easier for the teachers to explain the texts through visual aids such as pictures and animation videos. In the interview, even one teacher supported this belief that:

*Visual aids including movies, video clips or powerpoint slides somehow enable them to get the gist of the literature texts.*

Overall, it can be interpreted that the majority of the teachers have positive perceptions of the use of visual aids in teaching literary texts. In other words, this is a clear indicator that the use of visual aids in literature teaching is relevant as it meets the needs of the second language learners in reading and learning literary texts. Besides, it is enjoyable to teach using visual aids which leads to the improvement in students' performance in literature.

## 4.3 The Utilization of Visual Aids

This section presents the findings regarding the teachers' perceptions on how visual aids can be fully utilized as an effective learning tool in motivating students to read literary texts. The majority (96.2 %) of the teachers believed that the use of visual aids creates enjoyable learning environment in the literature classroom. This was probably due to the characteristics of visual aids like sound, light and color that could trigger and stimulate students' interest in learning literary texts and thus automatically creates a fun learning environment. Most of the respondents also stated that visual aids can be fully utilized as an effective learning tool as they help students to understand clearly the literary concepts in the literary texts. This phenomenon was most probably because of the graphics used in the visual aids undeniably served as a helpful tool in facilitating students' understanding of the literary concepts occurred in the texts. Interestingly, this probability was supported by Pillai and Vengadasamy (2010) who claimed that the use of graphics, illustrations, pictures, audio, and video is a useful tool in assisting students' understanding of the literary concepts in the texts [11].

The majority (88.5 %) of the teachers had positive perceptions in response to 'the use of visual aids arouses students' understanding of the importance of literature'. This was probably because the teachers realized that literature helped their students in their personal growth, cultural enrichment and most importantly language development [8], [10] & [13].

Moreover, most (86.5%) of the respondents believed that the use of visual aids assists students to cope with the complexity of the language used in the literary texts. One of the interviewed teachers also commented that:

*Videos and pictures allow the students to comprehend better because they will be able to see what is exactly happening in the literary texts and they don't really rely on listening where they might lose their concentration.*

Similarly, most (86.5 %) of the respondents believed that the use of visual aids helps the students to cope with cultural elements embedded in the texts. This was probably because the use of audio visual, animated videos and films, graphic pictures and other visual aids helped the students to visualize the texts clearly. This probability is tallied with what Muller (2006) claimed that the use of film for instance, helps the students to visualize clearly the literary theory and cultural aspects embedded in the texts [12]. Moreover, the majority of the teachers indicated their agreement with 'the use of visual aids generates students' creative thinking skills'. This was probably because the teachers felt that various activities could be implemented in literature teaching and thus it could help in fostering their creative thinking. The literature also deals a lot with abstract ideas and thus it really needs the students to do a lot of critical analysis.

## 5 Conclusions

The findings of this study are not only useful for English teachers teaching in the schools, but also to the lecturers with related field, especially those who are teaching trainee teachers and students of English literature course in the teacher training colleges, universities and other educational institutes. The findings may be served as guidelines for teachers when implementing visual aids in teaching, as they want their students to fully concentrate on the lesson, by being aware of the expectations and needs in literature teaching. When the teachers know how to grab students' attention, teachers can provide a friendly and interesting atmosphere for the students to learn. This will encourage the students not to just learn by listening and writing what the teachers told, said and provided in the classroom, but they will find their own initiative to read what they learn in order to improve their own understanding towards the lesson. Furthermore, the implementation of visual aids in teaching literature is less time consuming. As a result, the teachers will have more ample time to create enjoyable classroom activities and conduct an effective teaching and learning process.